# SUMMARY OF THE WORKING GROUP ON 'SINGLE PARTICLE EFFECTS: PARASITIC LONG-RANGE BEAM-BEAM INTERACTIONS' *

V. Shiltsev[#], FNAL, Batavia, IL, U.S.A.; E. Métral, CERN, Geneva, Switzerland

## Abstract

There were three presentations given at the session "Single particle effects: parasitic long-range beam-beam interactions" [1, 2, 3] which were followed by discussions. Below we summarize major findings and discussions.

## TEVATRON AND LHC OBSERVATIONS

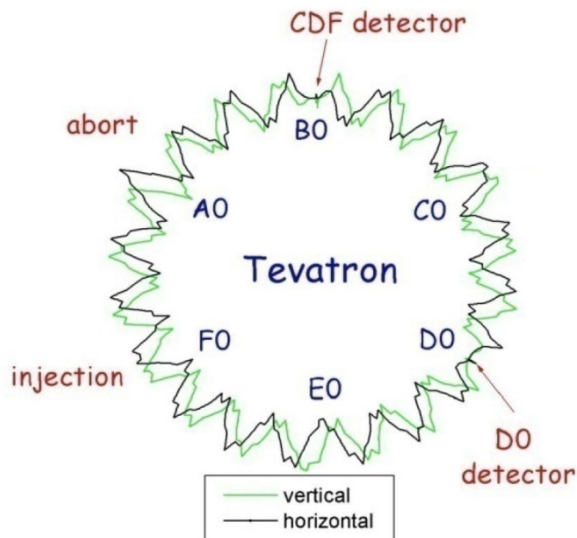

Figure 1: The pattern of the Tevatron helical orbits at the collision stage.

There are similarities and differences in the observations of the long-range beam-beam effects in the Tevatron and in the LHC. They start with the patterns of the parasitic interactions.

During the Tevatron Collider Run II 36 x 36 bunch operation, each bunch experienced 72 long-range interactions per revolution at injection, but at collision there were 70 long-range interactions and two head-on collisions per bunch at the CDF and D0 detectors (see Fig. 1). At the bunch spacing of 396 ns, the distance between the neighbor interaction points was 59 m. In total, there were 138 locations around the ring where beam-beam interactions occurred. The sequence of 72 interactions out of the 138 possible ones differed for each bunch, hence the effects varied from bunch to bunch. Notably, the long-range interactions occurred at the different betatron phases.

___

The locations of these interactions and the beam separations changed from injection to collision because of the antiproton cogging (relative timing between antiprotons and protons).

At the LHC, where the beams are separated with a crossing angle, there are up to 120 long range encounters which are lumped at the betatron phases of main interaction points (see Fig. 2). Consequently, the issues are very different from the helical (or pretzel) separation scheme.

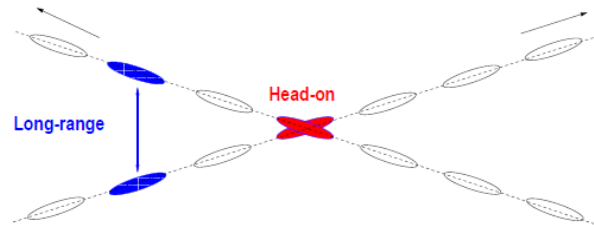

Figure 2: Schematic of proton-proton collisions in the LHC.

Besides the difference in the separation schemes and the total number of the parasitic interaction points, one should note that the LHC has larger separation – of about 9-10 $\sigma$ - in all interaction points, except one (LHC-b at the interaction point 8) where the separation varies during the collision runs from few to one $\sigma$ in order to level the luminosity at some 10% of the main low-beta interaction points (at CMS and ATLAS). During the Tevatron collision stores most of the long-range interactions were at 8-10 $\sigma$ but they were less essential than 4 near interaction point crossings at 5.8-6 $\sigma$ separation. In low-beta squeeze the beams briefly (2 s) came within 2-2.5 $\sigma$ at 1 parasitic interaction point and that usually caused sharp loss spikes. So, here the first unresolved question – why one such small-separation interaction point was so harmful in the Tevatron and seemingly is of no concern in the LHC? One can point to the difference in the single bunch intensities (1.2-1.5 $10^{11}$ protons per bunch in the LHC and some 3 $10^{11}$ protons per bunch in the Tevatron) but at this moment it is not clear whether that is sufficient for full explanation.

It was shown that in the Tevatron, the long-range beam-beam interactions occur at all stages (injection, ramp, squeeze, collisions) and affected both proton and antiproton beams. They resulted in beam losses, and emittance blow-ups, which occurred in remarkable bunch-to-bunch dependent patterns. Of notice is that these phenomena were a) thoroughly studied experimentally; b) described by phenomenological models indicating quantitative dependencies of the beam loss rates and the emittance growth rates on the machine and beam parameters (tunes, chromaticities, separations, beam intensities and emittances, etc); c) modeled in Lifetrac [4] simulations which not only described the observations but were used to make quantitative predictions (which were later confirmed in operation).

Studies of the beam-beam effects in the LHC are currently at the stage of compilation of the experimental evidences and analysis of parametric dependencies (on the crossing angle, intensities, tunes, bunch spacing, etc.). Collider operation and machine performance analysis tools are being developed, and the Tevatron SDA software (Software for Data Analysis) and on-line store analysis programs are being used as an example. The LHC beam diagnostic suite is being steadily expanding and improving with the goal of having several trustable, cross-calibrated monitors of all beam parameters working in bunch-by-bunch measurements modes.

Given detrimental consequences of the beam-beam effects (including long-range) on the Tevatron performance, the beam-beam issues have been seriously addressed and eventually corrected to the operational satisfaction. In particular, the long-range effects were mitigated by: i) an increase of separation by installation of additional HV separators; ii) a rearrangement of the helical orbits; iii) an optimization of the machine optics - linear and nonlinear; iv) pulsed e-lenses; v) a large number of incremental improvements (there was no "silver bullet"). In the LHC some of the most obvious operationally harmful beam-beam effects were corrected by proper adjustment of the beam loading schemes to equalize at least the number of the head-on collisions for all the bunches.

## SIMULATION OF LONG-RANGE AND HEAD-ON BEAM-BEAM EFFECTS

There are several approaches to the simulations of the beam-beam effects: i) the fastest is analytical calculations of the resonance driving terms [5] or similar method of calculating "smears" [3]; ii) fast tracking – by, e.g. Sixtrack [6] or frequency map analyses [7] – to find the dynamic aperture; iii) slow ("comprehensive") tracking of the long-term dynamics, e.g. with Lifetrac as described in [8]. The later method was shown to be very useful, adequate, having valuable quantitative predictive and provide results which can be directly compared to observables (lifetime, emittance growth, etc.). For instance, for most of the Collider Run II the modified Lifetrac weak-strong beam-beam code was used to study the beam-beam effects in the Tevatron. It correctly described all observed beam dynamics effects, had predictive power and had been particularly useful for supporting and planning changes of the machine configuration.

Methods i) and ii) are very practical and (relatively) very fast but their result – dynamic aperture – though potentially "measurable" in dedicated beam studies, does not provide quantitative description of the observables. Still, the dynamic aperture (DA) analysis is helpful as it gives qualitative estimates, e.g. the scaling laws for the LHC:

$$DA \propto \frac{1}{n_b}$$

$$DA \propto \frac{1}{\sqrt{\varepsilon}}$$

$$DA \propto d_{sep} \propto \alpha$$

$$DA \propto d_{sep} \propto \sqrt{\beta^*}$$

$$DA \propto \frac{1}{N}$$

where, $n_b$ is the number of bunches, $\varepsilon$ is the transverse emittance, $d_{sep}$ is the beam separation, $\alpha$ is the crossing angle, $\beta^*$ is the betatron function at the interaction point and $N$ is the bunch intensity.

## OTHER DISCUSSIONS

There also was an interesting discussion on the "complexity" of accelerators, understood in the mathematically defined terms of the *CPT theorem* [9]. At the very general level, it was pointed out – see [3] and Fig. 3 – that the hadron beam machines seems to be more "complex" (problematic, "not-that-easy to work with") than the electron ones; that the colliders are more "complex" than one beam machines; and that, seemingly, the most complex systems are those that involve more beams, e.g. 4-lepton-beams DCI collider [10], or 3-beam systems such as "beam-beam-beam" or "three beam instability" (two colliding hadron beams interacting with electron cloud) [11] or the beam-beam effects in hadron colliders compensated by electron lenses.

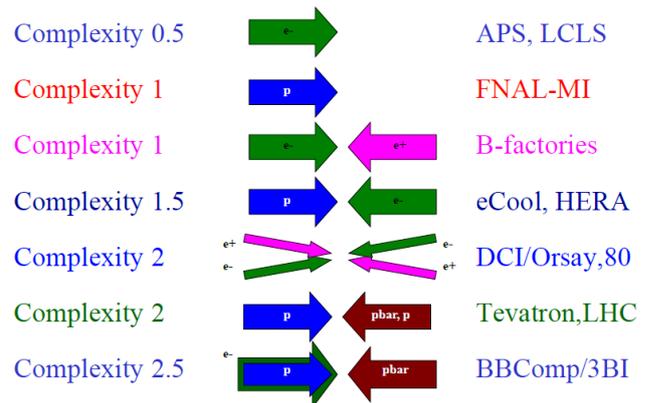

Figure 3: Simplified evaluation of the "complexity" of accelerators [3].

## REFERENCES

[1] W.Herr, these proceedings.
[2] D.Kaltchev, these proceedings.
[3] V.Shiltsev, these proceedings.

[4] D. Shatilov, Proceedings of the 2005 Particle Accelerator Conference, Knoxville, Tennessee, USA.
[5] Y. Alexahin, FERMILAB-TM-2148 (2001).
[6] F. Schmidt, CERN/SL/94–56 (AP).
[7] J. Laskar, Proceedings of the 2003 Particle Accelerator Conference, Portland, Oregon, USA.
[8] A. Valishev et al., JINST **7** (2012) P12002.
[9] V. Shiltsev, Mod.Phys.Lett. A, **26** (2011) 761.
[10] G. Arzelia et al., HEACC, 1971.
[11] A. Burov, these proceedings.